\documentstyle[prl,aps,preprint,tighten]{revtex}
\begin{document}
\draft
\title{Identity of the van der Waals Force and the Casimir Effect
and the Irrelevance of these Phenomena to Sonoluminescence}
\author{Iver Brevik\thanks{E-mail: iver.h.brevik@mtf.ntnu.no}}
\address{Division of Applied Mechanics, Norwegian University of
Science and Technology, N-7034 Trondheim, Norway}
\author{V. N. Marachevsky\thanks{E-mail: root@vm1485.spb.edu}}
\address{Department of Theoretical Physics, St. Petersburg University,
198 904 St. Petersburg, Russia}
\author{Kimball A. Milton\thanks{E-mail: milton@mail.nhn.ou.edu}}
\address{Department of Physics and Astronomy, The University of
Oklahoma, Norman, OK 73019 USA}
\preprint{OKHEP-98-08}
\date{\today}
\maketitle

\begin{abstract}
We show that the Casimir, or zero-point, energy of a dilute dielectric
ball, or of a spherical bubble in a dielectric medium, coincides with the sum of
the van der Waals energies between the molecules that make up the medium.
That energy, which is finite and repulsive
 when self-energy and surface effects are
removed, may be unambiguously calculated by either dimensional continuation
or by zeta function regularization.  This physical interpretation of the Casimir energy seems
unambiguous evidence that the bulk self-energy cannot be relevant to
sonoluminescence.
\end{abstract}
\pacs{78.60.Mq,42.50.Lc,12.20.Ds,03.70.+k}

\paragraph*{Introduction}
The Casimir effect has been recognized as a fundamental aspect of
quantum field theory for 50 years \cite{casimir}. This phenomenon, first
presented as an attractive force between parallel perfectly conducting
plates, may be thought of as  a result of changes in the electromagnetic
field fluctuations induced by the presence of boundaries.  Recently, it
has been confirmed to good accuracy by direct measurements \cite{lamoreaux,roy},
although the closely related Lifshitz theory \cite{lifshitz} was confirmed
experimentally 25 years ago \cite{anderson}.

Actually, the history of the effect goes back a bit further.  Casimir and
Polder had worked out the retarded dispersion force between molecules in
1947 \cite{cp}, the long range part of the van der Waals force.  Niels
Bohr shortly thereafter suggested to Casimir that zero-point energy was
relevant to the effect \cite{casatleipzig}, and subsequently Casimir presented
a derivation of the force between molecules, and between a molecule and a
conducting plate, based on such considerations \cite{casinparis}.
The derivation of the force between parallel plates followed shortly
\cite{casimir}.  It was thus clear from the outset that there was an intimate
tie between the van der Waals forces and the Casimir effect.

\paragraph*{Identity of van der Waals and Casimir forces}

The explicit demonstration of the identity of these two forces was given
in the case of dilute parallel dielectric slabs, where the Lifshitz formula 
for the Casimir energy may be easily seen to be equal, if dispersion be
neglected, to the sum of pairwise long-range van der Waals energies
\cite{lifshitz}:
\begin{equation}
V=-{23\alpha_1\alpha_2\over4\pi r^7},
\label{vdw}
\end{equation}
the Casimir-Polder retarded dispersion potential \cite{cp}. (See also 
Ref.~\cite{sdm}.)  Here the connection between dielectric constant $\epsilon$ 
and polarizability $\alpha$ is $\epsilon=1+4\pi N \alpha$,
$N$ being the number density of molecules.
The simple geometry makes this calculation easy and unambiguous.

The corresponding calculation for a spherical geometry is fraught with more
difficulty.  The sum of van der Waals interactions (\ref{vdw}) for a spherical
ball (the case for a spherical bubble is identical) has been given in 
Ref.~\cite{mng}.  A sensible procedure for carrying out the calculation
is dimensional continuation, which has been advocated in Ref.~\cite{benmil}.
That is, we evaluate the integral
\begin{equation}
E_{\rm vdW}=-{23\over 8\pi}\alpha^2N^2\int d^Dr d^Dr'(r^2+r^{\prime2}
-2rr'\cos\theta)^{-\gamma/2},
\label{evdw}
\end{equation}
by first regarding $D>\gamma$ so that the integral exists.  The integral
may be done exactly in terms of gamma functions, which when evaluated at
$D=3$, $\gamma=7$ yields \cite{mng}
\begin{equation}
E_{\rm vdW}={23\over1536\pi a}(\epsilon-1)^2.
\label{Evdw}
\end{equation}
(Note that the expression (\ref{evdw}) is formally negative or attractive,
while the continued result is positive or repulsive.)

Of course, the above calculation in three dimensions is divergent.  These
divergences are of two kinds: ``volume'' and ``surface.''  The volume divergence
is a self-energy effect that would be present if the medium filled all space,
and makes no reference to the interface, and therefore is quite unobservable.
If the divergences are regulated by inserting a point-splitting cutoff,
and the divergent terms are simply omitted, the same result (\ref{Evdw}) is
again obtained.

Now we turn to the Casimir effect.  For the case of the dielectric sphere
this was first worked out in Ref.~\cite{milton}.  That result has been
rederived, for the more general case of a spherical bubble, of radius $a$, 
having permittivity $\epsilon'$ and permeability $\mu'$, surrounded by
an infinite medium of permittivity $\epsilon$ and permeability $\mu$, in
Ref.~\cite{miltonng}.  Here the volume effect, corresponding to the intrinsic
self-energy of either medium, was explicitly removed;  a more detailed
justification of that procedure is given in Ref.~\cite{mng}.  The general
result is rather complicated:
\begin{eqnarray}
E_C&=&-{1\over4\pi a}\int_{-\infty}^\infty dy\, e^{iy\delta}\sum_{l=1}^\infty
(2l+1)\bigg\{x {d\over dx}\ln D_l+2x'[s_l'(x')e_l'(x')-e_l(x')s_l''(x')]
\nonumber\\
&&\mbox{}-2x[s_l'(x)e_l(x)-e_l(x)s_l''(x)]\bigg\},
\label{casimir}
\end{eqnarray}
where
\begin{equation}
D_l=[s_l(x')e_l'(x)-s_l'(x')e_l(x)]^2-\xi^2[s_l(x')e_l'(x)+s_l'(x')e_l(x)]^2,
\end{equation}
which uses the abbreviation
\begin{equation}
\xi={\left({\epsilon'\over\epsilon}{\mu\over\mu'}\right)^{1/2}-1\over
\left({\epsilon'\over\epsilon}{\mu\over\mu'}\right)^{1/2}+1}.
\end{equation}
The integration variables are $x=\sqrt{\mu\epsilon}|y|$, $x'=
\sqrt{\mu'\epsilon'} x$,
and the Ricatti-Bessel functions are
\begin{equation}
s_l(x)=\left(\pi x\over2\right)^{1/2}I_{l+1/2}(x),\quad
e_l(x)=\left(2x\over\pi\right)^{1/2}K_{l+1/2}(x).
\end{equation}  The formula (\ref{casimir}) has been regulated by a
time-splitting parameter, $\delta=\tau/a\to0$, where $\tau$ is a Euclidean
time separation between field points.

The general expression (\ref{casimir}) is rather opaque.  Therefore, we
consider a dilute dielectric ball, which was already considered in 
Ref.~\cite{milton}. (That is, we consider $\mu=1$ everywhere, and $\epsilon=1$
outside of the ball.)
 The formula, which still admits of dispersion, becomes
in that case
\begin{eqnarray}
E_C\approx-{1\over8\pi a}\sum_{l=1}^\infty (2l+1){1\over2}\int_{-\infty}^\infty
dy\, e^{iy\delta}(\epsilon(y)-1)^2 x{d\over dx}F_l(x),
\label{smalleps}
\end{eqnarray}
where
\begin{equation}
F_l(x)=x^2\left(1+{l(l+1)\over x^2}
\right)-{1\over4}\left({d\over dx}e_ls_l\right)^2-x^2\left[2\left(1
+{l(l+1)\over x^2}\right)e_ls_l-{1\over2}{d^2\over dx^2}e_ls_l\right]^2.
\end{equation}
The integrand here may be approximated by the uniform asymptotic
approximation \cite{abrom}:
\begin{equation}
e_l(x)s_l(x)\sim {1\over2}zt\left(1+{a_1(t)\over \nu^2}+{a_2(t)\over\nu^4}
+\dots\right),
\end{equation}
where $\nu=l+1/2$, $x=\nu z$, and $t=(1+z^2)^{-1/2}$.  The coefficients
$a_k(t)$ are polynomials in $t$ of degree $3k$.
If we ignore dispersion, and set the time-splitting parameter $\delta=0$,
we obtain \cite{brevik}
the leading uniform asymptotic approximation to (\ref{smalleps}),
\begin{equation}
E_C\sim {(\epsilon-1)^2\over 64 a}\sum_{l=1}^\infty\left\{\nu^2-{65\over128}
+{927\over16384\nu^2}+O(\nu^{-4})\right\}.
\label{asymp}
\end{equation}
The first two terms are formally divergent, but may be evaluated by
the zeta-function definition,
$\sum_{l=0}^\infty \nu^{-s}=(2^s-1)\zeta(s)$.
Note that if only the leading term were kept, the result given in
Ref.~\cite{miltonng} would be obtained, 
$E_1\sim -(\epsilon-1)^2/(256 a)$,
while including two terms reverses the sign and hardly changes the
magnitude \cite{brevik}:
$E_2\sim +33(\epsilon-1)^2/(8192 a)$.
This would seem to resolve the conundrum found in Ref.~\cite{miltonng},
the apparent sign disparity between the Casimir effect and the van der
Waals energy.

Indeed, let us do the result exactly.  (Probably it is possible to do
the integrals analytically, but we have not immediately seen how to do this.)
We simply add and subtract the two leading asymptotic terms from the
integrand in (\ref{smalleps}), so that $E=E_2+E_R$, where the
remainder is
\begin{equation}
E_R={(\epsilon-1)^2\over4\pi a}\sum_{l=1}^\infty \nu^2\int_0^\infty dz
\left[F_l(\nu z)-{t^4\over4}+{t^{10}\over8\nu^2}(1+8z^2-5z^4+z^6)\right],
\end{equation}
According to the third term in Eq.~(\ref{asymp}),
the $z$ integral here is asymptotic to $927 \pi/262144 \nu^4$; we evaluate
the $l$ sum by doing the integral numerically for the first ten terms,
and using the asymptotic approximant thereafter. The result is
\begin{equation}
E_C=(\epsilon-1)^2{0.004767\over a}.
\label{Ecas}
\end{equation}
This precisely agrees with the van der Waals result (\ref{Evdw}).\footnote{
As this paper was being written, we received a manuscript from G. Barton
\cite{barton}, in which exactly the same identity between the van der
Waals and Casimir forces was noted.  He uses an elementary method of
summing zero-point energies directly in powers of $(\epsilon-1)$, using
ordinary perturbation theory. Of course, our approach is in principle more
general, in that it allows for arbitrary $\epsilon$ and permits inclusion
of dispersion.} (The approximation $E_2$ is 15\% too low,
whereas if the first three terms in Eq.~(\ref{asymp}) are kept, the estimate
is 1.8\% high.)

\paragraph*{The Irrelevance to Sonoluminescence}
There has recently been considerable controversy concerning the possible
relevance of the Casimir effect to sonoluminescence
\cite{sonorev}.  The idea that
the ``dynamical Casimir effect'' might be relevant to sonoluminescence
originated in the work of Schwinger \cite{js}.  Recently, a series
of papers has strongly advocated Schwinger's point of view 
\cite{carlson,visser}.  This view has been criticized elsewhere \cite{mng}.
However, now that we clearly see that the Casimir energy may be
identified with van der Waals interactions, it seems perfectly plain that
the volume effect they consider, proportional to $\epsilon-1$, simply
cannot be present, because such cannot arise from pairwise interactions.
(This point was already made in Ref.~\cite{milton}.)
Our interpretation stands vindicated: an effect proportional to the volume
represents a contribution to the mass density of the material, and cannot
give rise to observable effects.  (A direct refutation of the photon
production calculation of Ref.~\cite{visser} will appear elsewhere 
\cite{leipzig}.)

More subtle is the role of surface divergences \cite{miltonng}.  The zeta
function regularization calculation we presented above simply discards such
terms; but they appear in more physical regularization schemes.  For
example, if the time splitting parameter in Eq.~(\ref{casimir}) is retained,
we get from the leading asymptotic expansion
\begin{equation}
E_{\rm div}=-{(\epsilon-1)^2\over4a}{1\over\delta^3};
\end{equation}
and if a simple model for dispersion is used,
with characteristic frequency $\omega_0$, the same result is obtained
with $1/\delta\to\omega_0a/4$ \cite{miltonng}. (A very similar result
is given in Ref.~\cite{brevik}.)
We believe these terms are probably also unobservable, for they modify
the surface tension of the liquid, which, like the bulk energy, is
already phenomenologically described.  (That surface tension has its origin
in the Casimir effect was proposed in Ref.~\cite{sdm}.)
In any case, this surface energy
is probably too small, and definitely of the wrong sign, to be relevant
to sonoluminescence.  (The flash of light is emitted at the minimum
radius.)

We note that Barton in his recent work \cite{barton} seems to concur
with our assessment: The terms ``proportional to $V$ [the volume]
 and to $S$ [the surface] would
be combined with other contributions to the bulk and to the surface
energies of the material, and play no further role if one uses the
measured values.''

It is truly remarkable that however the (true) divergences in the theory
are regulated, and subsequently discarded, the finite result is unchanged.
That is, in the van der Waals energy, we can simply omit the point-split
divergences, or proceed through dimensional continuation, where no divergences
are explicit; in either case, the same result (\ref{Evdw}) is obtained.
Likewise, the same result is obtained for the Casimir energy using either
a temporal point-splitting, or an exponential wavenumber cutoff \cite{barton},
and omitting the divergent terms; or through the formal trick of zeta-function
regularization.  It is worth re-emphasizing that we are not claiming that
the Casimir effect for a dielectric ball is finite, unlike the classic
case of a spherical conducting shell \cite{shell}.  It is merely that those
divergent terms serve to renormalize phenomenological parameters in the condensed
matter system.

So, finally, we are left with the finite term, which in the dilute
approximation is given by (\ref{Evdw}) or (\ref{Ecas}).  For a bubble
of minimum radius $\sim10^{-4}$ cm, the corresponding Casimir energy is
only $E_C\sim 10^{-3}$ eV.  This is ten orders of
magnitude to small to be relevant to sonoluminescence, where about one
million optical photons are emitted per flash, and again the sign is
wrong.  (As to the relevance of a static calculation to the dynamical
regime of sonoluminescence, we note that the adiabatic approximation
seems favorable, since the time scale for the flash, $\sim 10^{-11}$ s
is far longer than the time scale for optical photons, $\sim 10^{-15}$ s.)

\paragraph*{Acknowledgments}
We are indebted to Gabriel Barton for discussions in Leipzig, and for
sharing a draft of his paper with us. We are grateful to H. B. G.
Casimir and H. Rechenberg for conversations on the history of the Casimir
effect.  We thank Michael Bordag for
discussions, and for his organization of the Fourth Workshop on Quantum
Field Theory Under External Conditions in Leipzig, which allowed this
collaboration to take place.  This work was supported in part by a grant
from the U.S. Department of Energy.


\begin{references}
\bibitem{casimir} H. B. G. Casimir, Proc. K. Ned. Akad. Wet. {\bf 51}, 793 
(1948).

\bibitem{lamoreaux} S. K. Lamoreaux, Phys. Rev. Lett. {\bf 78}, 5 (1997).

\bibitem{roy} U. Mohideen and A. Roy, ``A Precision Measurement of the
Casimir Force from 0.1 to 0.9 Microns,'' physics/9805038.

\bibitem{lifshitz} E. M. Lifshitz, Zh.\ Eksp.\ Teor.\ Fiz.\ {\bf 29},
894 (1955) [English translation: Soviet Phys.\ JETP {\bf 2}, 73 (1956)];
I. D. Dzyaloshinskii, E. M. Lifshitz, and L. P. Pitaevskii, Usp.\ Fiz.\
Nauk {\bf 73}, 381 (1961) [English translation: Soviet Phys.\ Usp.\
{\bf 4}, 153 (1961)]; L. D. Landau and E. M. Lifshitz, {\it Electrodynamics
of Continuous Media\/} (Pergamon, Oxford, 1960), pp.\ 368--376.

\bibitem{anderson} E. S. Sabisky and C. H. Anderson, Phys.\ Rev.\ A
{\bf 7}, 790 (1973).

\bibitem{cp} H. B. G. Casimir and D. Polder, Phys.\ Rev.\ {\bf 73},
360 (1948).

\bibitem{casatleipzig} H. B. G. Casimir, talk at the Fourth Workshop on
Quantum Field Theory Under the Influence of External Conditions, Leipzig,
September 14, 1998.

\bibitem{casinparis} H. B. G. Casimir, talk at Colloque sur la theorie
de la liaison chemique, Paris, April 12--17, 1948.

\bibitem{sdm} J. Schwinger, L. L. DeRaad, Jr., and K. A. Milton,
Ann. Phys. (N.Y.), {\bf 115}, 1 (1978).

\bibitem{mng} K. A. Milton and Y. J. Ng, Phys. Rev. E {\bf 57}, 5504 (1998).


\bibitem{benmil} C. M. Bender and K. A. Milton, Phys. Rev. D {\bf 50}, 6547
(1994); K. A. Milton, {\it ibid.} {\bf 55}, 4940 (1997).

\bibitem{milton} K. A. Milton, Ann. Phys. (N.Y.) {\bf 127}, 49 (1980).

\bibitem{miltonng} K. A. Milton and Y. J. Ng, Phys. Rev. E {\bf 55}, 4207 
(1997).

\bibitem{abrom} M. Abramowitz and I. A. Stegun, {\it Handbook of Mathematical
Functions\/} (National Bureau of Standards, Washington, 1964), p.~378.

\bibitem{brevik} I. Brevik and V. N. Marachevsky, ``Casimir Surface Force
on a Dilute Dielectric Ball.''


\bibitem{barton} G. Barton, ``Perturbative Check on the Casimir Energy
on a Nondispersive Dielectric Ball.''

\bibitem{sonorev} B. P. Barber, R. A. Hiller, R. L\"ofstedt, S. J.
Putterman, and K. Weniger, Phys.\ Rep.\ {\bf 281}, 65 (1997).

\bibitem{js} J. Schwinger, Proc.\ Natl.\ Acad.\ Sci.\ USA {\bf 90},
958, 2105, 4505, 7285 (1993); {\bf 91}, 6473 (1994).

\bibitem{carlson} C. E. Carlson, C. Molina-Par\'\i s, J.
P\'erez-Mercader, and
M. Visser, Phys.\ Lett.\ {\bf B395}, 76 (1997); Phys. Rev. D {\bf 56},
1262 (1997);
C. Molina-Par\'\i s and M. Visser,  Phys. Rev. D {\bf 56}, 6629 (1997).  
Apparently, this work grew out of the suggestion that Gamma Ray Bursters have
their origin in the Casimir effect: C. Carlson, T. Goldman, and J.
P\'erez-Mercader,
Europhys. Lett. {\bf 36}, 637 (1996).

\bibitem{visser} S. Liberati, M. Visser, F. Belgiorno, and D. W. Sciama.
``Sonoluminescence: Bogolubov Coefficients for the QED Vacuum of a
Collapsing Bubble,'' quant-ph/9805023; S. Liberati, F. Belgiorno, 
M. Visser, and D. W. Sciama, ``Sonoluminescence as a Quantum Vacuum
Effect,'' quant-ph/9805031.

\bibitem{leipzig} K. Milton, talk at the Fourth Workshop on
Quantum Field Theory Under the Influence of External Conditions, Leipzig,
September 14--18, 1998.

\bibitem{shell} T. H. Boyer, Phys.\ Rev.\ {\bf 174}, 1764 (1968);
B. Davies, J. Math.\ Phys.\ {\bf 13}, 1324 (1972);
R. Balian and B. Duplantier, Ann.\ Phys.\ (N.Y.)
{\bf 112}, 165 (1978);
K. A. Milton, L. L. DeRaad, Jr., and J. Schwinger,
Ann.\ Phys.\ (N.Y.) {\bf 115}, 388 (1978).

\end{references}
\end{document}